\documentclass[final,1p,times]{elsarticle}

\usepackage{amssymb}

\journal{Physica A}

\begin{document}

\begin{frontmatter}



\title{Clausius inequality and H-theorems\\ for some models of random wealth exchange}


\author[Lebedev,DP]{Sergey M. Apenko}

\address[Lebedev]{ P N Lebedev Physical Institute, Leninsky pr. 53, Moscow, 119991, Russia}
\address[DP]{Russian Endowment for Science and Education, Moscow, Russia}
\ead{apenko@lpi.ru}

\begin{abstract}
We discuss a possibility of deriving an $H$-theorem for nonlinear discrete time evolution equation that describes random wealth exchanges. In such kinetic models economical agents exchange wealth in pairwise collisions just as particles in a gas exchange their energy. It appears useful to reformulate the problem and represent the dynamics as a combination of two processes. The first is a linear transformation of a two-particle distribution function during the act of exchange while the second one  corresponds to new random pairing of agents and plays a role of some kind of feedback control. This representation leads to a Clausius-type inequality which suggests a new interpretation of the exchange process as an irreversible relaxation due to a contact with a reservoir of a special type. Only in some special cases when equilibrium distribution is exactly a gamma distribution, this inequality results in the $H$-theorem with monotonically growing `entropy' functional which differs from the Boltzmann entropy by an additional term. But for arbitrary exchange rule the evolution has some features of relaxation to a non-equilibrium steady state and it is still unclear if any general $H$-theorem could exist.

\end{abstract}


\end{frontmatter}


\section{Introduction}
\label{sec:1}

Models of random wealth exchange between economical agents are still rather popular in econophysics though they have been studied already for more than several decades (see e.g. reviews \cite{Y,rev,reva} and references therein). These models are used to study equilibrium distributions of wealth and treat systems of agents like a gas, where particles exchange their energy in pairwise collisions. Probably one of the most general models of this kind is a model with savings \cite{sav}. In this model we first randomly pair agents which have some (positive) amounts of wealth $x$ and $y$ and after the exchange these values are modified $x,y\rightarrow x',y'$ where
\begin{equation}\label{one}
x'=\lambda x+(1-\lambda)\epsilon (x+y), \qquad y'=\lambda y+(1-\lambda)(1-\epsilon)(x+y),
\end{equation}
and $\lambda,\epsilon\in[0,1]$ are some dimensionless parameters, which in general may be random.  Interpretation of these equations is rather obvious: agents first make savings, i.e. they lay aside a fraction $\lambda$ of their money and then go to the market with the rest of it. Then in each act of interaction (collision) between two of them they simply randomly redistribute between themselves the total amount of money they both have brought. Clearly the total wealth is conserved in such an exchange, $x+y=x'+y'$, so this is actually just a kind of redistribution of wealth. Then the process is repeated many times until finally the wealth distribution function $p(x)$ tends to some equilibrium one $p_0(x)$.

When this model was first introduced, the saving propensity $\lambda$ was taken as a fixed  parameter while $\epsilon$ was a random number uniformly distributed between 0 and 1 \cite{sav}. Numerically investigated equilibrium distribution was found to be very close to a gamma distribution $p_0(x)\sim x^a\exp(-bx)$ with $a=3\lambda/(1-\lambda)$  and $b=(a+1)/\langle x\rangle$ determined by the mean wealth $\langle x\rangle$ \cite{savg}. Later it was shown however  that the exact equilibrium solution clearly deviates from gamma distribution (see e.g. \cite{nogamma}), though the latter is still a surprisingly good approximation. The model was then studied further and its different modifications were proposed, where $\lambda$ could be random and inhomogeneous (see e.g. \cite{sava} and reviews \cite{Y,rev,reva}). In what follows we will consider the general case when both $\lambda$ and $\epsilon$ are random numbers with some probability distributions $P(\lambda)$ and $D(\epsilon)$. We will also assume that agents are indistinguishable, so that $D(\epsilon)=D(1-\epsilon)$  and hence the mean value of $\epsilon$ is just $1/2$.

It is interesting that even before these models appeared in economical context  mappings like (\ref{one}) were discussed by mathematicians as a special kind of  `smoothing transformations'  \cite{pot,smth}, but these papers remained practically unknown to physicists. In these papers general existence theorems were proved and even certain equilibrium solutions  $p_0(x)$ were found. The special case $\lambda=0$ corresponds to  Dr\u{a}gulescu-Yakovenko model \cite{DY}, but is also known as Ulam's redistribution of energy problem \cite{U}. For absolutely random redistribution in Ulam's problem, when  $D(\epsilon)=1$, it was shown that in equilibrium $p_0(x)=b\exp(-b x)$  \cite{DY,U}, where $1/b=\langle x\rangle$ is determined by the mean energy $\langle x\rangle$. Equilibrium solution was also found for a more general case when $D(\epsilon)$ is given by a symmetric beta distribution $D(\epsilon)\sim\epsilon^{a}(1-\epsilon)^{a}$ with arbitrary $a>-1$ and is exactly the gamma distribution \cite{smth} (see also \cite{T}).

From the very beginning it was clear that for a general exchange with $\lambda\neq 0$ equilibrium distribution is not of a simple Gibbs-like  form $\sim\exp(-bx)$ and hence it does not correspond to the maximum of the Boltzmann entropy under the constraint $\langle x\rangle=\rm{const}$. Even for $\lambda=0$ if e.g. during the exchange the total wealth is divided into two equal parts ($D(\epsilon)$ is a delta-function at $\epsilon=1/2$) all agents will finally have the same amount of wealth   \cite{U}, which corresponds to an ordered state with minimum entropy. Kinetic equation approach to exchange models also indicates the absence of the classical Boltzmann $H$-theorem in general case \cite{kinet}. Such a behavior was qualitatively explained by a presence of some kind of Maxwell's demon in the system, related to the breakdown of the time-reversal symmetry at microscopic level \cite{Y}. It seems that such processes may look not quite `natural' and the system of agents is probably better regarded as an open one with some additional external agent, or device, that could divide different amounts of money  in a given ratio and then prepare new random pairs. Thus we arrive at a general problem of how the second law should look like in models of random exchanges when some control agent similar to Maxwell's demon may be present. 

To study this problem we need first to specify in more detail how the exchange process goes in time. Normally this is not needed because in econophysics we are primarily interested in final equilibrium wealth distribution, but now we certainly have to chose some dynamics of exchanges. In what follows we will use statistical physics language i.e. call agents `particles' and wealth `energy' and chose simple synchronous dynamics  used e.g. in \cite{U} and in a series of papers by L\'{o}pez-Ruiz and his colleagues \cite{gl}. That is, we will assume that all particles from a vast set are first paired, then all collisions take place simultaneously in all pairs and after that a new pairing is performed. This implies a discrete time evolution, where at each time step the energy distribution transforms as $p(x)\rightarrow p'(x)$ with
\begin{eqnarray}\label{gl}
p'(x)=\int_0^{\infty}\int_0^{\infty}dudv\,
\langle\delta\left[x-\lambda u-(1-\lambda)\epsilon(u+v)\right]\rangle_{\lambda,\epsilon} p(u)p(v)=\nonumber\\
=\int_0^{\infty}\int_0^{\infty}dudv \,\frac{\theta(u+v-x)}{u+v}W(x; u,v)p(u)p(v).
\end{eqnarray}
where $\langle\ldots\rangle_{\lambda,\epsilon}$ is the average over the distributions $P(\lambda)$ and $D(\epsilon)$. This is a kind of a nonlinear mapping which conserves the mean energy  $\langle x\rangle$.  In the integrand we have the probability density $p(u)p(v)$ that particles in a given pair have energies $u$ and $v$ multiplied by the probability of the transition $u,v\rightarrow x$. We have also extracted the factor $1/(u+v)$ from this probability and introduce a new  kernel $W(x;u,v)$ which is a homogeneous function of its arguments, i.e. $W(xc;uc,vc)=W(x;u,v)$ for arbitrary $c$. This property is rather obvious because $W$ is dimensionless and the exchange rule (\ref{one}) does not contain any dimensionful parameters except for energies of colliding particles. Theta-function in (\ref{gl}) explicitly ensures that $u+v>x$. For $\lambda=0$ and uniform redistribution law we have $W=1$ and only the factor $1/(u+v)$  remains \cite{gl}, which is just the normalized probability distribution for $x$ that can have any value between zero and $u+v$. 

Nonlinear mapping (\ref{gl}) is in fact a simple example of what is sometimes called  a `nonlinear Markov process' \cite{NM}, because it looks like an ordinary Markov chain with transition probabilities depending on distribution function itself. Another interesting analogy is with the renormalization group (RG) transformation. Indeed, suppose that we have a one-dimensional chain of non-interacting continuous `spins' of  magnitude $x$. If we now introduce blocks of two adjacent spins of lengths $u$ and $v$ and define $x$ as a new block variable, then  (\ref{gl}) may be viewed just as a usual real space block-spin RG transformation \cite{ren} of the probability distribution. Stationary distributions $p_0(x)$ then corresponds to fixed points of such an RG transformation.

Since this evolution is based on pairwise interactions, just as in a gas of particles, described by the Boltzmann equation, one might ask whether some kind of a modified $H$-theorem could still be valid here. But despite the simplicity of the model it is not so easy to find the answer because usual methods, like e.g. those used in \cite{kinet}, are not directly applicable to discrete time evolution. Moreover, contrary to the continuous time Boltzmann equation the time scale here is different. In a gas we normally expect monotonicity theorems to be valid on a time scale which is much larger than the time between collisions. But for our synchronous dynamics we are looking for monotonicity at each step of discrete time, i.e. on a microscopic time scale. Next,  perturbation due to collisions is in no sense small, since all collisions take place simultaneously and after each act of redistribution $p(x)$ changes significantly. 

Despite these arguments it was shown that at least for $\lambda=0$ and for uniform redistribution law $D(\epsilon)=1$ the Boltzmann entropy
\begin{equation}
 S_B(p)=-\int dx\,p(x)\ln p(x)
\end{equation}
still increases at each time step \cite{gl,A}, i.e. $S_B(p')\geq S_B(p)$. However for general exchange models  a question about the proper $H$-theorem (if it exists)  remained open.

To study this question we  first generalize the approach of \cite{A} which makes use of a two-particle distribution function $f(x,y)$. This formalism is briefly reviewed in Sec. 2. Each step of nonlinear evolution is represented as a combination of two steps: the first one is a linear transformation which describes the change in $f(x,y)$ due to scatterings within random pairs of particles, while in the second step the two-particle function reduces to a factorized form as a result of new pairings. In Sec. 3 we derive a general inequality, similar to Clausius inequality in equilibrium thermodynamics, for the change in entropy at each step. This inequality suggests that `Maxwell's demon' that controls exchanges may be viewed as simply a special kind of external reservoir with some `heat' flow to it during the exchange process.  However, this Clausius-type inequality leads to $H$-theorems only in some special cases, when $p_0(x)$  in equilibrium  is exactly a gamma distribution. This takes place e.g. for Ulam's problem with beta redistribution law \cite{smth,T} and for directed random market model \cite{dir}. These models are discussed in Sec. 4  where for both cases a monotone functional $S(p)$ is obtained which differs from the Boltzmann entropy by an additional term. We also briefly discuss some possible interpretations of this functional.

\section{Two-stage representation of the exchange process}
\label{sec:2}

Our search for the $H$-theorem is based on the approach of \cite{A} which transforms initial nonlinear problem into a linear one, supplemented by a subsequent projection operation, by introducing a two particle distribution function $f(x,y)$. After a collision and energy exchange $f(x,y)\rightarrow f'(x,y)$ and it is possible to write down  a simple equation describing this evolution.

There exists a regular way to obtain an equation for the two-particle probability distribution function $f(x,y)$ from the original equation for $p(x)$. We start from the integral representation for the theta-function in (\ref{gl})
\begin{equation}\label{theta}
\theta(u+v-x)=\int_0^{\infty}dy\,\delta(x+y-u-v).
\end{equation}
Since the delta-function here obviously represents energy conservation law, it is quite natural to view $y$ as energy of the second particle after the collision. Now we substitute this expression into eq. (\ref{gl}) and after changing the order of integration we may write
\begin{equation}\label{mar}
p'(x)=\int_0^{\infty}dy\,f'(x,y),
\end{equation}
where $f'(x,y)$ is given by
\begin{equation}\label{evf}
f'(x,y)=\int_0^{\infty}\int_0^{\infty}dudv\,\frac{\delta(x+y-u-v)}{u+v}W(x;u,v)f(u,v),
\end{equation}
with $f(u,v)=p(u)p(v)$. 

Physically it is rather obvious that $f(x,y)$ is a two-point distribution function i.e. the probability distribution that one particle has energy $x$ while the other has $y$. Then (\ref{evf}) shows how this two-particle distribution transforms as a result of the collision. It should be noted, however, that Eq. (\ref{evf}) alone does not describe correctly the true evolution of the two-particle probability distribution in exchange problem. It takes into account only collisions within fixed pairs of particles while the true evolution includes also new random pairing of particles at each step, not accounted for in (\ref{evf}). 

The transformation of the original nonlinear equation made above suggests a new representation of the problem. In terms of of the two-particle function $f(x,y)$ the evolution may be represented as consisting of two steps. 
\begin{enumerate}
\item At the first step, when we pair particles at random, we have initially $f(x,y)=p(x)p(y)$ and then as a result of collisions
\begin{equation}
f(x,y)=p(x)p(y)\rightarrow f'(x,y),
\end{equation}
where the new probability distribution is given by (\ref{evf}). 
\item Next we perform a new random pairing, destroying all correlations induced by interaction. Therefore this  second step may be describes as
\begin{equation}\label{red}
f'(x,y)\rightarrow p'(x)p'(y),
\end{equation}
where $p'(x)$ is the marginal distribution given by (\ref{mar}). One can then easily verify that combination of these two steps is completely equivalent to the original nonlinear equation.
\end{enumerate}
The procedure described above is, in a sense, opposite to how the Boltzmann equation is derived from the equation for the  many-particle distribution function, since here we in fact restore the true two-particle evolution from the Boltzmann-like equation. And the factorized initial condition $f(u,v)=p(u)p(v)$  is actually just the well known `molecular chaos' assumption (see, e.g.  \cite{kac}). Our inverse problem is, however, more simple, since we already have this Boltzmann-like equation with time-reversal symmetry explicitly broken, so this procedure is exact and rather straightforward.

The second step of dynamics to some extent resembles some measurement process or feedback control. But contrary to existing examples of stochastic processes with measurements (see e.g. \cite{mes}) here we do not have any particular quantity that is measured. Probably it is better to say that here the entire single-particle distribution function $p(x)$  is measured after the collisions to produce new initial data for the next step, and such a feedback control seems quite different from what is usually studied.

\section{Energy exchange as  interaction with environment. Clausius inequality}
\label{sec:3}

It is useful to make now a transformation to  new variables, which are more natural for the exchange process (\ref{one}). Let us introduce instead of $x$ and $y$ the total energy of a given pair $E$ and a fraction $\xi<1$ of this total energy that one of the particles has
\begin{equation}\label{nvar}
E=x+y,\qquad \xi=\frac{x}{x+y}.
\end{equation}
In terms of these variables the exchange process (\ref{one}) looks very simple: 
$$
\xi\rightarrow\lambda\xi+\epsilon(1-\lambda),
$$
while $E$ remains unchanged. The two-particle distribution function in terms of these new variables looks as
\begin{equation}\label{fnew}
\phi (E,\xi)=Ef(\xi E,(1-\xi)E),
\end{equation}
where the additional factor $E$ appears because of the normalization condition (it cancels the Jacobian of the transformation from $\xi,E$ to $x,y$).

Let us now rewrite our evolution equation (\ref{evf}) in terms of the function $\phi(E,\xi)$. For this reason we  make a change of variables in the integral $u,v\rightarrow E',\xi '$ where $E'=u+v$ and $\xi '=u/(u+v)$. The Jacobian of this transformation is just $u+v$ so it exactly cancels the denominator in the integrand. Integration over $E'$ is trivial because of the delta-function, so finally we arrive at
\begin{eqnarray}\label{lin}
  \phi'(E,\xi)=\int_0^1 d\xi '\,W[\xi E;\xi ' E,(1-\xi ')E]\phi(E,\xi ')=\nonumber\\
=\int_0^1 d\xi '\,W[\xi ;\xi ',1-\xi ']\phi(E,\xi '),
\end{eqnarray}
because of the  homogeneity of the function $W(x;u,v)$. This transformation conserves
positivity of $\phi(E,\xi)$ and its normalization because $\int d\xi W[\xi ;\xi ',1-\xi ']=1$.  It is clear also that the entire marginal distribution of $E$ given by $\int d\xi\phi(E,\xi)$  is not changed during collisions.

The advantage of Eq. (\ref{lin}) is that this is a linear equation, just a sort of a stationary Markov chain, for which the monotone functional can be constructed in a regular way. Normally, it is the relative entropy, or Kullback-Leibler distance \cite{cover} to equilibrium,
\begin{equation}\label{r}
K=\int_0^{1} d\xi\, \phi(E,\xi)\ln\frac{\phi(E,\xi)}{\phi_0(\xi)},
\end{equation}
where $\phi_0(\xi)$ is a stationary solution of eq. (\ref{lin}). In what follows we will assume that such a stationary solution does exist, is strictly positive and can be taken as a function of only $\xi$, because the kernel $W$ in (\ref{lin}) does not depend on $E$.  Monotonicity here follows from the chain rule for the relative entropy \cite{cover} and means that
\begin{equation}\label{rel}
\int_0^{1} d\xi\, \phi(E,\xi)\ln\frac{ \phi(E,\xi)}{ \phi_0(\xi)}\geq
\int_0^{1} d\xi\, \phi'(E,\xi)\ln\frac{ \phi'(E,\xi)}{ \phi_0(\xi)},
\end{equation}
which can be also rewritten as
\begin{eqnarray}
\int_0^{1} d\xi\, \phi(E,\xi)\ln\phi(E,\xi)-\int_0^{1} d\xi\, \phi'(E,\xi)\ln\phi'(E,\xi)\geq \nonumber\\
\geq\int_0^{1} d\xi\,[\phi(E,\xi)-\phi'(E,\xi)]\ln\phi_0(\xi)
\end{eqnarray}
We can now integrate this inequality over $E$ and from the relation (\ref{fnew}) it follows that on the left hand side we have just the difference of Boltzmann entropies $S_B(f')-S_B(f)$, where
\begin{equation}
S_B(f)=-\int dxdy\,f(x,y)\ln f(x,y)
\end{equation}
Next, recall that initial distribution is factorized, $f(x,y)=p(x)p(y)$, and hence the initial entropy equals
\begin{equation}\label{init}
S_B(f)=2S_B(p).
\end{equation}
And for the final distribution we have a simple information theory inequality 
\begin{equation}\label{reduct}
2S_B(p')\geq S_B(f'),
\end{equation}
which follows from the positivity of the mutual information \cite{cover}
\begin{equation}
 I= \int\int dxdy\,f'(x,y)\ln\frac{f'(x,y)}{p'(x)p'(y)}=2S_B(p')-S_B(f')\geq 0
\end{equation} 
of $x$ and $y$ variables after collisions. Physically  inequality (\ref{reduct})  means that for a partition of a system into two parts the total entropy is smaller than the sum of entropies of subsystems. Combining all these formulas we finally arrive at the inequality
\begin{eqnarray}\label{main}
 S_B(p')-S_B(p)\geq \frac{1}{2}\int_0^{1} d\xi\int_0^{\infty}dE\,[\phi(E,\xi)-\phi'(E,\xi)]\ln\phi_0(\xi).
\end{eqnarray}

In the case when $\phi_0(\xi)$ is constant, which is the case e.g. for the uniform redistribution of energy at $\lambda=0$, the right hand side vanishes and we have just usual entropy growth $S_B(p')\geq S_B(p)$ \cite{gl,A}. Unfortunately, in general $\phi'(E,\xi)$ cannot be expressed in terms of $p'$, and so we cannot derive any suitable $H$-theorem from Eq. (\ref{main}). But we will see now that actually this inequality may be interpreted as Clausius inequality for an irreversible process, accompanied by some `heat' flow.

Indeed, since we assume $\phi_0(\xi)$ to be strictly positive we can represent it in exponential form and introduce a new `pseudoenergy' ${\cal E}(\xi)$ for pairs of particles according to
\begin{equation}\label{gibbs}
\phi_0(\xi)=\frac{1}{Z}\exp(-{\cal E}(\xi)/T),\qquad Z=\int_0^1d\xi \exp(-{\cal E}(\xi)/T)
\end{equation}
Of course, introducing an independent `temperature' $T$ does not make much sense, it is done just to preserve the common formulas. However, in the next section we will see that sometimes this temperature may still have some meaning. Using (\ref{gibbs}) our main inequality (\ref{main}) may be rewritten as
\begin{equation}\label{cla}
S_B(p')-S_B(p)\geq\frac{Q}{T},
\end{equation}
where the transferred `heat' $Q$  is defined as a change of mean pseudoenergy (per particle) during collisions,
\begin{equation}\label{q}
Q=\frac{1}{2}\int_0^1 d\xi {\cal E}(\xi)[\phi'(\xi)-\phi(\xi)],
\end{equation}
where $\phi(\xi)$ and $\phi'(\xi)$ are marginal distributions, obtained from $\phi(E,\xi)$ before and after collisions by integration over $E$.

Inequality (\ref{cla}) has now the form of Clausius inequality from ordinary equilibrium thermodynamics (for non-equilibrium extensions of Clausius theorem see e.g. \cite{clau}). This  inequality suggests a new interpretation of the exchange process. We should probably consider our system of particles as an open one, which at every step of evolution is brought in contact with some external  reservoir  in a very specific way. Namely, we pair particles at random and then the distribution in $\xi$  irreversibly relaxes towards an equilibrium with the environment while the amount of `heat' $Q$ is transmitted to the system. The peculiarity of this approach is that the system is coupled to the reservoir not through the exchange of the true energy $x$, as is normally the case, but  pseudoenergy ${\cal E}(\xi)$,  defined only for pairs of particles. 

It is important to note here that in general the stationary solution $\phi_0(\xi)$  of equation (\ref{lin}) actually may not be reached, because the relaxation is always interrupted by new pairings. This means that in general we may have $\phi(\xi)\neq\phi'(\xi)$ even if  $p(x)=p'(x)=p_0(x)$. Then $Q$ is nonzero even in a stationary state and is negative, as follows from (\ref{cla}) for $S_B(p)=S_B(p')$, i.e. there is a constant heat flow to the reservoir. But this implies in its turn that in the two-stage representation we have rather a non-equilibrium steady state instead of a true equilibrium, even if $p(x)$ tends to a single stationary distribution.  To some extent the resulting state resembles the one found in \cite{saad}, where equilibrium domains were observed in non-equilibrium systems. Relaxation of $p(x)$ looks like a simple approach to equilibrium $p_0(x)$ while the evolution of the two-particle function $f(x,y)$  in such a steady state is in fact a cyclic process, consisting of relaxation towards the equilibrium $\phi_0(\xi)$ followed by the subsequent  new random pairing that drives pairs away from this equilibrium. It is not clear whether in such a situation any $H$-theorem may exist for the `subsystem' described by $p(x)$.

\section{$H$-theorems for Ulam's redistribution problem and directed random market model}
\label{sec:4}

There exist, however, special cases, when $Q=0$ in a stationary state. The first example is a special case  of exchange with $\lambda=0$, also known as Ulam's redistribution of energy problem \cite{U}, with $D(\epsilon)$ given by beta distribution \cite{smth,T}, while the second one corresponds to the so called directed random market model \cite{dir}. 

\subsection{Ulam's redistribution of energy problem}

In this case $\lambda=0$ and we have only a redistribution of the total energy of a given pair 
\begin{equation}\label{ulam}
x'=\epsilon(x+y),\qquad y'=(1-\epsilon)(x+y)
\end{equation}
according to some probability distribution $D(\epsilon)$. This  is in fact just the Dr\u{a}gulescu-Yakovenko model \cite{DY} but with arbitrary redistribution law. Then for the kernel in the original  evolution equation we have
\begin{equation}
W(x;u,v)=(u+v)\langle\delta[x-\epsilon(u+v)]\rangle_{\epsilon}=D\left(\frac{x}{u+v}\right).
\end{equation}

Linear equation (\ref{lin}) for the two-particle distribution function now has a very simple form
\begin{equation}\label{nlin}
\phi '(\xi,E)=D(\xi)\int_0^1 d\xi ' \phi(\xi ',E)
\end{equation}
This means that during the collision all information about the original distribution in $\xi$ is completely erased and the distribution in question is simply substituted by $D(\xi)$. The stationary solution of (\ref{nlin})  may be taken as
\begin{equation}
\phi_0(\xi)=D(\xi)
\end{equation}
and we can treat the process (\ref{nlin}) as instantaneous relaxation to thermal equilibrium. 

An interesting and important case corresponds to a symmetric beta distribution
\begin{equation}\label{beta}
D(\epsilon)=C\epsilon^{a}(1-\epsilon)^{a}, \quad \epsilon\in [0,1]
\end{equation}
where $C$ is a normalization constant and $a>-1$ is a parameter that determines the shape of the distribution. When $a\rightarrow\infty$ this distribution tends to a delta-function located at $\epsilon=1/2$, i.e. each particle acquires exactly one half of the total energy, while for negative $a$ the distribution $D(\epsilon)$ diverges at $\epsilon=0$ and $\epsilon=1$ indicating that after the collision one particle normally gets much more energy than the other.

For these redistribution laws equilibrium solutions are already known to be gamma distributions
\begin{equation}\label{equ}
p_0(x)\sim x^{a}\exp(-bx)
\end{equation}
with $b$ determined by the conserved mean energy. These solutions were obtained for the pure gambling model of Bassetti and Toscani \cite{T}, which is in fact a continuous time version of Ulam's redistribution problem (see also \cite{pot}). 

If we now introduce pseudoenergy ${\cal E}(\xi)$ according to $D(\xi)\sim\exp(-{\cal E}(\xi)/T)$  with $D(\xi)$ from (\ref{beta}) we may interpret $a$ as inverse temperature, i.e. take $1/T=a$ while ${\cal E}(\xi)=-\ln\xi(1-\xi)$. Such an interpretation indeed makes sense because e.g. the case $a\rightarrow\infty$ then corresponds to zero temperature and so it is quite natural that in this case equilibrium corresponds to a completely ordered state while infinite $T$ results in $a=0$ and uniform redistribution law.

It is important that for $p_0(x)$ given by gamma distribution (\ref{equ}) the equilibrium two-particle distribution function, which before the collision is factorized, $f_0(x,y)=p_0(x)p_0(y)$,  is factorized also in variables $\xi$ and $E$. Indeed, in this case
\begin{eqnarray}\label{phiz}
\phi_0(\xi,E)=Ep_0(\xi E)p_0((1-\xi)E)\sim\nonumber\\
\sim\xi^{a}(1-\xi)^{a} E^{2a+1}\exp(- bE)
\end{eqnarray}
is simply a product of $D(\xi)$ and a function that depends only on $E$. According to (\ref{nlin}) this factorization in its turn implies that {\em during the collision the two-particle distribution function in this case is not changed and hence remains factorized also after the collision}, so that $f_0'(x,y)=f_0(x,y)=p_0(x)p_0(y)$ is a stationary solution of the linear evolution equation (\ref{evf}). This shows how peculiar gamma distribution is and this observation is close, in a sense, to an interesting recent demonstration of how equilibrium gamma distributions naturally arise in case of factorization  of the many-body distributions \cite{prad}.

For $D$ given by the symmetric beta distribution $\ln\phi_0(\xi)$  is just $a\ln\xi(1-\xi)$  up to an unimportant constant. Then changing variables in the expression for the transferred heat $Q$  from $E,\xi$  back to $x=\xi E$ and $y=(1-\xi)E$ we can write
\begin{equation}
\ln\xi(1-\xi)=\ln x+\ln y-2\ln(x+y)
\end{equation}
and
\begin{eqnarray}\label{heat}
Q/T=\frac{1}{2}\int_0^1 d\xi \int_0^{\infty}dE[\phi(E,\xi)-\phi'(E,\xi)] \ln \phi_0(\xi)=\nonumber\\
=\frac{a}{2}\int_0^{\infty}\int_0^{\infty}  dxdy\, [f(x,y)-f'(x,y)](\ln x+\ln y)=\\
=a\int_0^{\infty} dx\, [p(x)-p'(x)]\ln x\nonumber
\end{eqnarray}
since terms $\sim\ln(x+y)$ cancel because the total energy of a pair $x+y$ and its entire distribution are not changed under energy redistribution.

But then Clausius inequality obviously can be rewritten as a sort of $H$-theorem 
$$
S(p')\geq S(p)
$$
 with
\begin{equation}\label{fina}
S(p)=S_B(p)+a\int_0^{\infty}dx\,p(x)\ln x.
\end{equation}
Only for the uniform redistribution law, when $a=0$, it is the Boltzmann entropy $S_B(p)$  that always grows. This new $S$-functional is maximized by the equilibrium distribution $p_0(x)$ from (\ref{equ}) under the constraint that the mean energy is fixed.

$S$-functional (\ref{fina}) has a very general form, the same as discussed e.g. by Attard \cite{At}. Entropy usually has such a form when $x$ represents some macrostate for which a lot of microstates are possible, so that a macrostate has nonzero entropy. In our case this implies that the entropy of a given `macrostate', labeled by $x$, is equal to $a\ln x$. In fact the total entropy of a large system (typically consisting of some subsystem and a reservoir) always can be represented in a similar general form  when expressed in terms of a distribution function $p(x)$ for a subsystem. 

The $S$-functional obtained may be also written as
\begin{equation}\label{fin}
S(p)=-\int_0^{\infty}dx\,p(x)\ln \frac{p(x)}{x^a}.
\end{equation}
and is also of the same type as the entropy recently proposed for random mixture problems \cite{et}. In their paper Maynar and Trizac argued that because of the measure problem the entropy for a continuous variable should generally look like $-\int dx\,p(x)\ln[\Lambda(x)p(x)]$ and clearly (\ref{fin}) is exactly the same with $\Lambda(x)=x^{-a}$.  It should be noted, however, that this result was derived in \cite{et} from the Jacobian of the transformation $x,y\rightarrow x',y'$ while in our case the Jacobian of (\ref{ulam}) is exactly zero (for $\lambda=0$ original transformation does not have the inverse) so that the method of Maynar and Trizac does not directly work here.

The representation (\ref{fin}) also suggests that probably $S$ may be interpreted as just the usual entropy, but for some multidimensional problem. Indeed, suppose that in $d$-dimensional gas there is a velocity distribution $\varphi({\bf v})$. If $\varphi$ depends only on the absolute value $x=|{\bf v}|$  we may introduce a new function $p(x)\sim x^{d-1}\varphi(x)$, normalized as $\int dx\, p(x)=1$, and the Boltzmann entropy equals
\begin{eqnarray}
S=-\int d{\bf v}\,\varphi({\bf v})\ln\varphi({\bf v})=
 -\sigma_d\int_0^{\infty} dx\, x^{d-1}\varphi(x)\ln\varphi(x)=\nonumber \\
=-\int_0^{\infty} dx\,p(x)\ln\frac{p(x)}{x^{d-1}}+\rm const
\end{eqnarray}
which is the same as (\ref{fin}) up to a constant. So probably redistribution problem with $a\neq 0$ is a projection of some yet unknown $a+1$-dimensional problem with uniform redistribution of some vector quantity. This in fact follows from the observation that any gamma distribution may be viewed as energy distribution in a multidimensional gas, which was noticed before in many papers devoted to  wealth exchange models \cite{rev}. 

It is interesting that because of energy conservation $S$-functional from (\ref{fin}) is (up to a constant) just minus the relative entropy $\int p(x)\ln p(x)/p_0(x)$ with $p_0$ from (\ref{equ}). So the relative entropy monotonically decreases just as for a stationary  Markov process with linear master equation, despite the nonlinearity of the original equation for $p(x)$. Similar observation concerning relative entropy behavior was made recently for a nonlinear Boltzmann equation with non-conservative interactions \cite{B}. It is not clear, however, how such monotonicity could be obtained analytically directly from the original equation (\ref{gl}) for $p(x)$. 

\subsection{Directed random market model}
Directed random market model was introduced by  Mart{\'i}nez-Mart{\'i}nez  and L\'{o}pez-Ruiz \cite{dir} to study the situation when after a collision one of the agents may acquire much more `energy'  than the other. In this case they have proposed a mapping 
\begin{equation}
x'=\lambda x,\qquad y'=y+(1-\lambda)x
\end{equation}
with random $\lambda$ uniformly distributed on [0,1], $P(\lambda)=1$. Since agents are indistinguishable we have to assume that for  $x$ and $y$ interchanged the same mapping is valid. Then it is easy to verify that this kind of exchange is actually just the special case of the general model (\ref{one}) when $\epsilon$ is either 0 or 1  with equal probability.

This model is in fact a particular case of Angle's model \cite{ang}, while for fixed  non-random $\lambda$ it would be just the multiplicative asset exchange model of Ispolatov, Krapivsky and Redner \cite{IKR}.

Evolution equation for this model was derived in \cite{dir} 
\begin{equation}\label{dirmar}
p'(x)=\frac{1}{2}\int_x^{\infty}du\,\frac{p(u)}{u}+
\frac{1}{2}\int_0^x dv\int_{x-v}^{\infty}du\,\frac{p(u)p(v)}{u}
\end{equation}
and it is easy to check that it may be represented in the same general form as before (\ref{gl}) with
\begin{equation}
W(x;u,v)=\frac{u+v}{2}\left(\frac{\theta(u-x)}{u}+\frac{\theta(x-v)}{u}\right)\,.
\end{equation}
This means that the kernel in the integral equation (\ref{lin}) for the two-particle distribution function equals
\begin{equation}
W(\xi;\xi',1-\xi')=\frac{\theta(\xi'-\xi)}{2\xi'}+\frac{\theta(\xi-1+\xi')}{2\xi'}\,.
\end{equation}
Then the linear evolution equation for $\phi(E,\xi)$ may be written as
\begin{equation}\label{lindir}
\phi'(E,\xi)=\frac{1}{2}\int_{\xi}^{1} \frac{d\xi'}{\xi'}\phi(E,\xi')+
\frac{1}{2}\int_{1-\xi}^1\frac{d\xi'}{\xi'}\phi(E,\xi')
\end{equation}
and it has a normalized stationary solution
\begin{equation}
\phi_0(\xi)=\frac{1}{\pi}\frac{1}{\sqrt{\xi(1-\xi)}}\,.
\end{equation}
This solution is the same beta distribution as $D(\xi)$ in the previous example, but with $a=-1/2$. This is rather unexpected, because the exchange rule and both the nonlinear equation for $p(x)$ and linear equation for $\phi(E,\xi)$ are quite different. But since the stationary distribution $\phi_0(\xi)$ is the same, we can now simply repeat the previous calculation performed in (\ref{heat}) and obtain the same $H$-theorem with
\begin{equation}
S(p)=S_B(p)-\frac{1}{2}\int dx\, p(x)\ln x\,.
\end{equation}
This functional  grows at each step, $S(p')\geq S(p)$, and is maximized by $p_0(x)\sim x^{-1/2}\exp(-bx)$ for fixed mean energy. Clearly this $p_0(x)$ is the gamma distribution with $a=-1/2$ and one can verify by direct substitution that this is indeed the stationary solution of evolution equation (\ref{dirmar}). Thus though for many models with savings stationary distributions are known to be very close to a gamma distribution \cite{Y,rev,reva,savg}, for the directed market model this solution is exact. Quite recently this result was obtained also in a different way \cite{kat}.  For the general model with savings, however, even for fixed $\lambda$ (but random $\epsilon$) the linear equation for $\phi(E,\xi)$ is more complicated and no exact stationary solution is known.

\section{Discussion}
\label{sec:5}

We have tried to obtain an $H$-theorem for the model of random exchanges, but it appeared possible only for some special exchange rules, when equilibrium solutions are given by gamma distributions.  The relative entropy (Kullback-Leibler distance to equilibrium) for such exchanges decreases monotonically just as for an ordinary stationary  Markov chain, despite the nonlinearity of the initial evolution equation. Similar observation about relative entropy was made recently for systems with non-conservative interactions \cite{B}. Numerical analysis performed there suggests that monotonicity of the relative entropy may be quite a general feature of some non-linear evolution equations even though no general proof exists in this case. Unfortunately, for exchange models with arbitrary distributions of $\lambda$ and $\epsilon$ it appeared possible to derive analytically only a Clausius-type inequality (\ref{cla}) from which no general $H$-theorem follows.

In cases when $H$-theorem does exist, the monotonically growing $S$-functional differs from the usual Boltzmann entropy by an additional term and can be interpreted in different ways. Certainly, it is tempting to call this functional `entropy' in agreement with \cite{At,et}. One can also recall that for a system of hard spheres described by the nonlinear Enskog equation, the true entropy functional also differs from the simple Boltzmann entropy \cite{ens}. Indeed, it is generally accepted that Boltzmann's expression for the entropy is actually valid only for dilute gases while in less trivial situations the entropy functional should have some more complicated form.

However, in the present case it seems that quite a different interpretation is possible. Probably here there is no need to change the expression for entropy, but it is better to consider the system of particles as an open one. This means that we assume a presence of some additional agent, similar to Maxwell's demon (as was already noticed in \cite{Y}), that performs a controlling function during the energy exchange. While Maxwell's demon is usually associated with  information processing and erasure \cite{max}, here it seems more natural to interpret its action in terms of a `heat' flow to some external reservoir, as the Clausius inequality (\ref{cla}) suggests.

The two-stage evolution approach reveals also an unexpected peculiarity of the relaxation in exchange models. If we consider only the evolution of the single particle distribution function $p(x)$ then it looks like an ordinary relaxation to equilibrium. But in terms of the two-particle distribution $f(x,y)$ the situation is more interesting.  

For a general exchange rule the final state of evolution in this picture is not a true equilibrium, but rather some sort of a non-equilibrium steady state. In such a state $f(x,y)$ oscillates with constant production of `heat' (\ref{q}) in each act of energy exchange. This happens because the factorization of $f(x,y)$ at the beginning of each time step due to previous random pairing is destroyed by the following exchange, which introduces correlations between particles. In terms of kinetic theory of gases this means that `molecular chaos' condition \cite{kac},  in this model does not propagate in time and only the next pairing returns $f(x,y)$ back to its initial factorized form. Note that stationary energy distribution $p_0(x)$ remains unchanged during these oscillations of $f(x,y)$. 

This situation is close, in a sense, to appearance of equilibrium-like domains in non-equilibrium systems \cite{saad}. It is not clear whether in such cases one can expect any kind of $H$-theorem to be valid for a relaxing subsystem, which is a part of a larger system approaching non-equilibrium steady state. Indeed, it is not easy to obtain a proper state dependent Lyapunov functional even for simple Markov processes in the absence of detailed balance \cite{ness}. So probably there is no surprise that only for $p_0(x)$ given by gamma distributions when we have equilibrium also for the two-particle evolution it appears possible to derive the $H$-theorem. Whether some $H$-theorem could exist for general exchange models still remains an open problem.

\section*{Acknowledgments}

I am very grateful to A. Chakraborti, J. Gaite,  R. L\'{o}pez-Ruiz,  V. Losyakov for valuable discussions and to A. Puglisi and E. Trizac for stimulating correspondence. The work was supported in part by RFBR Grants No. 12-02-00520, 13-02-00457.

\end{document}